# ROTATING WINDS FROM ACCRETION DISKS IN CATACLYSMIC VARIABLES: ECLIPSE MODELING OF V347 PUPPIS[1]

Isaac Shlosman

*Department of Physics and Astronomy*

*University of Kentucky*

*Lexington, KY 40506-0055*

*E-mail: shlosman@asta.pa.uky.edu*

Peter Vitello

*Physics Department*

*Lawrence Livermore National Laboratory*

*L-14, P.O. Box 808, Livermore, CA 94550*

*E-mail: vitello@spaniel.llnl.gov*

and

Christopher W. Mauche

*Physics Department*

*Lawrence Livermore National Laboratory*

*L-41, P.O. Box 808, Livermore, CA 94550*

*E-mail: mauche@cygnus.llnl.gov*

[1] Based in part on observations with the *International Ultraviolet Explorer* satellite, which is sponsored and operated by the National Aeronautics and Space Administration, the Science Research Council of the United Kingdom, and the European Space Agency.

## ABSTRACT

We study the eclipsing nova-like variable V347 Pup by matching its UV emission line profiles in and out of eclipse to synthetic lines using a 3D kinematic and radiation transfer model. Our results support the accretion disk origin of winds in non-magnetic cataclysmic variables as opposite to the white dwarf origin. Our main point concerns the importance of rotation for the UV emission line shapes in such systems. In particular, we show that the narrowing of the UV emission lines in V347 Pup during eclipse can be easily explained by the eclipse of the innermost part of the wind by the secondary and the resulting reduction in the contribution of rotational broadening to the width of the lines. During the eclipse, the residual line flux is very sensitive to the maximal temperature of disk radiation. Good fits for reasonable mass-loss rates have been obtained for maximum disk temperatures $\approx$ 50,000 °K. This constraint was imposed either by leveling off the inner disk temperature profiles, in agreement with recent observations of some nova-like objects, or by assuming that the accretion disk does not extend to the surface of the white dwarf, in which case V347 Pup would be an intermediate polar. In anticipation of high-speed spectrophotometry of cataclysmic variables by the *Hubble Space Telescope*, we provide a numerical model of a time-resolved eclipse of V347 Pup or similar such system to be verified by future observations.

## 1. INTRODUCTION

Nova-like variables and dwarf novae in outburst are two types of non-magnetic cataclysmic variable (CV) stars which exhibit P Cygni line profiles in the ultraviolet. This phenomena is attributed to the high-velocity winds believed to exist in these systems (see Córdova 1994 for a recent review). While photospheric emission from the accretion disks in CVs is expected to dominate the UV continuum and to contribute to low ionization lines such as H$\alpha$ (e.g., Shakura & Sunyaev 1973; Mayo et al. 1980), the lines of highly ionized species, e.g., C IV $\lambda$1549, Si IV $\lambda$1398, and N V $\lambda$1240, form in the extended region above the disk. The strongest indication that this is the case comes from eclipse studies of high inclination angle CVs, e.g., UX UMa, RW Tri, OY Car, Z Cha, V Sge, and others. The amplitude of the variation in the lines of highly ionized species is much smaller than the comparable changes in the UV-optical continuum in and out of eclipse (Holm et al. 1982; King et al. 1983; Córdova & Mason 1985; Drew & Verbunt 1985; Williams et al. 1986; Naylor et al. 1988; Harlaftis et al. 1992a, b; Mauche et al. 1994).

Two type of wind models in non-magnetic CVs has been invoked so far. The first model assumes radial winds which emanate from the white dwarf that are ionized by the boundary layer and the disk radiation (Drew & Verbunt 1985; Kallman & Jensen 1985; Drew 1987; Mauche & Raymond 1987). The second model emphasizes winds from the extended surface of the accretion disk, and is characterized by inherently biconical symmetry supported by underlying ionizing continuum and the driving force in the wind (Shlosman, Vitello, & Shaviv 1985; Vitello & Shlosman 1988). Comparison of observed UV line profiles in CVs with synthetic lines originating in radial and disk winds exposes the problems associated with radial winds. First, radial winds require mass-loss rates similar to the accretion rates, or the wind will be over-ionized due to the geometrical dilution factor $\propto r^{-2}$, i.e., disk wind can be much more collimated. High mass loss rates create an energy problem if the wind is driven radiatively (e.g., Mauche & Raymond 1987; Shlosman & Vitello 1993, hereafter Paper I; Vitello & Shlosman 1993, Paper II). Second, radial winds have excessive absorption at large inclination angles (Paper I) unless they are moderately biconical (Drew 1987) and/or the disk is limb-darkened (Mauche & Raymond 1987). Most importantly, radial winds also ignore rotation which introduces multiple-scattering surfaces in the wind and a radial shear which affects the line optical depth and the resulting line core intensity (Paper I). As we show in the following, rotational broadening is essential to explain the observed eclipsing line profiles and, in fact, invalidates the hypothesis of winds centered on the white dwarfs.

The driving force behind the winds in CVs is most probably due to radiation pressure in the UV lines (Shlosman, Vitello & Shaviv 1985; Vitello & Shlosman 1988; Paper I), but contribution from hydromagnetic torques (Blandford & Payne 1982; Pudritz & Norman 1985) cannot be entirely excluded at present. Within this context, Emmering, Blandford & Shlosman (1992) presented a self-consistent 3D dynamic model of an hydromagnetically accelerating emission-line gas flow ('beads on a wire') from accretion disks in active galactic nuclei, which may be of relevance to CV winds as well. Either mechanism will lead to a dominant biconical symmetry in the wind with a superimposed orbital modulation resulting from departures in the axisymmetry of the radiation or magnetic fields. The biconical symmetry is confirmed by the sensitivity of the high-ionization line shapes to the inclination angle $i$ of the system. The absorption part of the P Cygni line profiles disappears for $i \gtrsim 65°$ and the emission lines appear broad and symmetric.

Eclipsing CVs are promising systems in which to study the wind geometry and kinematics. Among the eclipsing systems with available datasets, we focus on the nova-like variable V347 Pup, thereby building on the recent results of Mauche et al. (1994, hereafter Paper III). Observations of this CV in and out of eclipse reveal that the C IV, Si IV, and N V line profiles narrow substantially during eclipse. The same behavior is observed in OY Car (Drew 1991) and in UX UMa (Mason et al. 1995), and is contrary to that expected if these lines are formed in a radial wind. During the eclipse by the secondary of a radial wind, the line core is drastically reduced due to the complete eclipse of the low-velocity component of the wind near the white dwarf. The wings of the line are affected much less by the eclipse of the extended high-velocity component of the wind far from the white dwarf. As a result, the intensity of the line core falls faster than in the wings, leading to a broadening of the line profile during eclipse (Paper III). We will explain the observed narrowing of the UV line profiles during eclipse within the framework of *rotating* disk winds. We show, by means of numerical modeling of line formation in a 3D wind, that the eclipse of such a wind leads naturally to a decrease in the width of the emission line thus confirming our prediction to the importance of rotation for UV lines in CVs (Paper I).

Section 2 is devoted to an analysis of observational data on V347 Pup. Results of numerical modeling of UV emission lines in and out of eclipse are provided in Section 3 and their implications on the overall geometry of system and the wind are discussed in Section 4. We also present numerical modeling of a time-resolved synthetic eclipse for this CV to be verified by high-resolution observations with the *Hubble Space Telescope* (*HST*), which start to become available (e.g., Mason et al. 1995 for UX UMa). Conclusions are given in Section 5.

## 2. SPECIFICS OF V347 PUP SYSTEM

V347 Pup (which is also know as LB 1800) is a bright nova-like CV and the (purported) optical counterpart of the hard X-ray transient 4U 0608−49 (Buckley et al. 1990, but see Paper III). Buckley et al. derived the main parameters of the system. Its period is 5.56 hr, based on radial velocity and photometric variations. It is an almost totally eclipsing system with an inclination angle of $i = 87° \pm 3°$ and a mass ratio of $q = 0.46 \pm 0.04$. The He II $\lambda4686$ line appears flat-bottomed during the eclipse (Paper III). The radii of the secondary and the accretion disk are $\approx 4 \times 10^{10}$ cm. The ratio of these radii to the orbital separation is $0.31 \pm 0.01$. From the empirical mass-period relation (e.g., Patterson 1984) the secondary mass is found to be $\approx 0.55$ $M_\odot$. This brings the white dwarf mass to the unusually high value among eclipsing CVs of $M_{\rm WD} \approx 1.2$ $M_\odot$. The radius corresponding to this mass is $R_{\rm WD} = 3.9 \times 10^8$ cm (Hamada & Salpeter 1961). Otherwise, V347 Pup is similar in many respects to SW Sex and other eclipsing nova-like variables (Williams 1989) which have a typical accretion rates $\dot{M}_a \approx 10^{-8}$ $M_\odot$ yr$^{-1}$. For V347 Pup parameters, this rate cannot be smaller than $\approx 6$–$9 \times 10^{-9}$ $M_\odot$ yr$^{-1}$, or the disk is expected to experience dwarf nova outbursts (Shafter, Wheeler & Canizzo 1986; Paper III), which are not detected in this system. Mass-transfer rate of $10^{-8}$ $M_\odot$ yr$^{-1}$ is also consistent with the empirical relationship $\dot{M}_a$–Period found by Patterson (1984) for this period range. The eclipses observed in V347 Pup are deep ($\approx 3$ mag in $U$, slightly less in $V$) and last approximately 1 hr. Because the widths of the eclipses of the continuum and emission lines are comparable, the optical emission line regions should have a radial extent similar to the optical continuum region. Because of the system's high inclination, the soft X-ray flux observed by *ROSAT* during its all-sky survey is unlikely to be direct boundary layer emission. Finally, the distance to V347 Pup was estimated at 174–380 pc by Buckley et al. and $\approx 300$ pc by Paper III.

A reduction of a factor $\approx 2$ during eclipse was observed for the C IV line, $\approx 3$ for the N V, Si IV and Mg II lines, and $\approx 5$ for the He II $\lambda 1640$ line (Paper III). Also, the Gaussian widths of the C III, N V, Si IV and C IV emission lines were reduced by $\approx 30\%$ compared to their mean values out of eclipse, while the widths of the He II $\lambda 1640$ and Mg II lines remain unchanged or increased slightly. Also, the Gaussian width of the C IV line decreased from $\sigma \approx 5.0$ Å out of eclipse to $\sigma \approx 3.5$ Å in eclipse, but the line shapes appeared to be largely unaffected. Finally, the concave shape of the continuum with a minimum around 2400 Å hints towards possible contributions in the UV from two components with different temperatures, interpreted by Mauche et al. as the self-eclipsed accretion disk and the disk rim.

## 3. NUMERICAL MODELING OF LINE PROFILES

### 3.1. *Kinematics and Radiative Transfer*

Bright CVs are expected to transfer mass at a rate $\dot{M}_a \approx 10^{-8}$ $M_\odot$ yr$^{-1}$. Corresponding disk temperatures are high enough that much of the continuum is radiated in the UV. Conditions in the inner disk photosphere are similar to those found in OB star photospheres, leading to the possibility that high-velocity winds driven by radiation pressure in UV resonance lines may be the source of the CV wind (Shlosman, Vitello, & Shaviv 1985; Vitello & Shlosman 1988). Assuming that the UV line profiles are formed in the wind above the disk, we compare synthetic lines resulting from our modeling with observations. Comparison of profiles in and out of eclipse is used to constrain our numerical model.

We adopt the 3D kinematical and radiation transfer model of a line-driven wind (LDW) from a rotating disk described in Paper I. The photoionization of the LDW is assumed to come primarily from the accretion disk, with negligible contributions from the boundary layer and the white dwarf. Significant boundary layer emission was found invariably to lead to over ionization of the wind unless unreasonably high mass-loss rates were used. Below we mention other necessary details of the model. For further details, see Paper I.

The LDW streamlines are assumed to be helices which lie on conical surfaces with a constant opening angle. Streamlines start at the disk surface and continue with a constant angle $\theta$ relative to the rotation axis of the disk. The velocity of the LDW is given in terms of its cylindrical components, $v_r$, $v_\phi$, and $v_z$. On the streamline, $v_r = v_l \sin\theta$ and $v_z = v_l \cos\theta$, where $v_l$, the velocity in the $rz$-plane, is assumed to be given by a power law function

$$v_l = v_0 + (v_\infty - v_0)\left[\frac{(l/R_v)^\alpha}{(l/R_v)^\alpha + 1}\right] , \tag{1}$$

where $l^2 = (r - r_0)^2 + z^2$ is the distance from the disk along the conical streamline surface, $v_0$ and $v_\infty$ are the initial and asymptotic wind velocities along the streamline, $R_v$ is the LDW acceleration scale height, and $\alpha$ is the power-law constant. In the following we use the velocity law for which the wind accelerates marginally faster than linearly (i.e., the best fit obtained in Papers I and II). The asymptotic velocity is assumed to scale with the local escape velocity at $r_0$ in the disk, $v_{\rm esc} = (2GM_{\rm WD}/r_0)^{1/2}$, and $v_0$ is taken to be 6 km s$^{-1}$. The LDW orbital velocity $v_\phi$ follows from the assumptions of Keplerian motion at the base of the wind and the subsequent conservation of angular momentum.

The radiation transfer is performed applying the Sobolev approximation (Sobolev 1957) modified for a three-dimensional geometry as described by Rybicki & Hummer (1978, 1983). The Sobolev approximation, as we use it, is an approximation to the optical depth and not to the global treatment of the radiation transfer. We are calculating the probability of scattering along a particular ray neglecting redistribution. A detailed comparison of the Sobolev method with other radiation transfer treatments was made by Bastian et al. (1980). This study found that the Sobolev approximation gives very accurate results for flows with high velocity gradients.

The spectrum in the neighborhood of the line was obtained by evaluating separately the scattered luminosity from the wind and the net unscattered luminosity from the disk. Absorption was modeled by tracing rays from the disk surface towards the observer and reducing the original intensity by $e^{-\tau_s}$, where $\tau_s$ is the Sobolev optical depth. Hence multiple resonant surfaces are treated in detail for the original continium coming from the disk and the white dwarf. The source function is calculated using the disconnected approximation (Marti & Nordlinger 1977). This method ignores the contribution from other resonant velocity surfaces to the source function in the wind, in particular those introduced by rotation. The disk is assumed to be optically thick as long as its temperature exceeds 8000 $^{\rm o}$K. The study of the topology of the resonant surfaces in Paper I showed that secondary scattering contributions do not have a major effect on the source function, and the eclipse of the wind by the disk is negligible for the high inclinations we are interested in.

The effect of the eclipse on the line profile was included in our transport model by taking the secondary to be a sphere placed in the plane of the disk at the orbital separation distance from the white dwarf. For a particular choice of orbital phase, we then neglect intensity contributions for all rays towards the observer from either the accretion disk or from the wind which pass through the secondary.

Recent work by Knigge, Woods & Drew (1995) using a Monte Carlo radiation transfer scheme found line profiles which resemble those in Paper I quite closely for winds with both rotation and radial outflow, but differ in the singular case of a pure rotation. The pure rotation case disagreement is limited to the very line center where Knigge et al. observe an absorption spike instead of the small emission spike of Paper I. Although this difference may be attributed to the dissimilar techniques used in treating radiation transfer, we note that Knigge et al. assumed constant ionization fractions in the wind. Such an assumption is not an unreasonable approximation for radially outflowing stellar winds, where both the radiation field and the density vary roughly as $1/r^2$ over the bulk of the line emission region. For winds from CV accretion disks, the geometric factors from both the radiation field and the wind lead to very rapid 2D variations in the ionization fraction that can strongly affect the shape of the emission/absorption lines. It is our experience that the critical wind parameters and even the domain occupied by the LDWs from disks depend in a complex non-linear fashion on the wind ionization. Because of these extreme sensitivity of the wind ionization balance to the UV radiation of the inner disk, models which *presume* the ionization structure in the LDW can err significantly in estimating the values of system parameters. In this sense, the LDWs from OB stars appear to be much simpler and robust solutions of equations of radiation hydrodynamics.

As noted above, *IUE* spectra of V347 Pup reveal that the C IV line intensity during eclipse is roughly 70% of its value out of eclipse. As the secondary radius is comparable to the disk radius, the strong line intensity during eclipse is indicative of a very extensive wind emission region surrounding the white dwarf. We found that the LDW ionization structure is very sensitive to the temperature profile in the disk. The continuum intensities used in our modeling are taken to be that of a blackbody at the effective temperature $T_{\rm eff}(r_0)$ (see below) corrected for limb darkening by a factor $\frac{1}{2}(1+\frac{3}{2}\cos i)$. By analogy with OB stars, we assume that the wind originates on the part of the disk where $2\times 10^4 \lesssim T_{\rm eff}(r_0) \lesssim 5\times 10^4$ °K (Abbott 1982; Vitello & Shlosman 1988), where the upper and lower temperature limits are defined as $T_{max}$ and $T_{min}$. The ionization in the LDW is calculated assuming local ionization equilibrium and a constant temperature of 20,000 °K. Details of the ionization balance treatment are given in MacGregor & Vitello 1982. Ionization equilibrium may break down close to asymptotic velocity where the the photoionization and radiative recombination timescales may become longer than the dynamic timescale, leading to the ionization fraction being “frozen”. We find that in this region of the flow a constant ionization fraction is naturally achieved from the ionization balance assumption due to both the wind density and ratiation field both scaling roughly as $1/r^2$ far from the disk. The assumption of constant electron temperature is not a bad approximation because the temperature in the wind is a slowly varying function compared to the density and radiation field. All ions of H, He, C, N and O are considered with solar abundances.

For a standard steady-state optically thick accretion disk, the effective temperature at the disk surface at radius $r_0$ for blackbody emission is given by (Shakura & Sunyaev 1973)

$$T_{\rm eff}^4(r_0) = \frac{3GM_{\rm WD}\dot{M}_a(r_0)}{8\pi\sigma r_0^3}\left[1-\left(\frac{R_{\rm WD}}{r_0}\right)^{1/2}\right] , \qquad (2)$$

where the mass-accretion rate $\dot{M}_a$ in general can vary with radius in the disk due to mass-loss. The local intensity $I_\nu(r_0)$ was approximated by its blackbody value $B_\nu[T_{\rm eff}(r_0)]$. The local spectrum of the disk likely differs from that of a blackbody (Shaviv & Wehrse 1991), but as the ionization balance in the wind depends on the integral of $I_\nu(r_0)$ over different radii, it is less sensitive to the details of the disk spectra.

### 3.2. *Eclipse: the Effect of Rotation on Synthetic Lines*

The strong dependence of the line profiles on the inclination angle in CVs argues for a biconical rather than spherically-symmetric wind geometry. Both observational and theoretical line profiles for large inclination angles show pure emission (Papers I and II). Paper I shows explicitly such a transition from P Cygni line profiles to pure emission as the inclination angle increases beyond $\approx 65°$ (see their Figure 5). For angles $\gtrsim 80°$, line shapes are more or less symmetric. A closer examination of the synthetic line profiles, however, reveals a weak asymmetry in the line core, close to the emission peak: this is caused by rotation which modifies the projected resonant velocity surfaces. Radial winds produce substantial absorption at the same wavelengths, in contradiction with observations. Generally, the effect of rotation at the base of the wind is to drastically modify the line optical depth by introducing radial shear. Artificially switching off the rotation in a biconical disk wind leads to an increase in the line core intensity with respect to continuum and to a narrowing of the line (Paper I).

Spherically symmetric radial winds (Drew 1987; Mauche & Raymond 1987) have difficulties in reproducing many properties of winds in CVs. In particular, for the same mass-loss rates as disk winds, radial winds are more strongly ionized which results in lower optical depths at large distances from the white dwarf and shallower absorption depths of the UV resonance lines. At large inclination angles, radial winds show too much absorption or produce lines which are too narrow, in direct conflict with observations (Paper II). Most importantly, the line profiles do not narrow in radial winds during the eclipse (Paper III).

In Paper II we found that disk winds are capable of providing good fits to the line profiles of three CVs. However, unique fits are not possible, given the many degrees of freedom in *any* kinematical model and the limited number of parameters characterizing the profiles. The fewest constraints were found for non-eclipsing systems at high inclination angles where no absorption is evident in the UV line spectra. Modeling the line profiles of eclipsing CVs in and out of eclipse imposes needed constraints for high inclination angle systems. Fitting the line profiles in and out of eclipse with the same set of parameters increases the constraints on eclipsing systems to the same level as for CVs of moderate inclination where both absorption and emission components of the C IV line can be fit.

To compare our synthetic line profiles to observations, we have used the eclipse and mean out of eclipse C IV line profiles of V347 Pup shown in Figure 20 of Paper III. We scale all spectra to the peak intensity of the line out of eclipse. The choice of fiducial model parameters for V347 Pup is given in Table 1. The following system parameters were considered to be fixed: the mass and radius of the white dwarf, the white dwarf and boundary layer temperatures, $T_{WD}$ and $T_{BL}$, the radius of the disk, $R_d$, the radius of the secondary, $R_S$, the orbital separation, $a$, and the inclination angle. Note that the white dwarf and boundary layer temperatures were chosen to be so low as to not contribute to the photoionization on the wind. All other model parameters (including the mimimum and maximum opening angles of the wind, $\theta_{\rm min}$ and $\theta_{\rm max}$) were optimized in fitting the observed spectra. The parameters which were most extensively varied to study the solution phase space were the accretion rate and the mass-loss rate in the LDW from the disk. In generating synthetic spectra in and out of eclipse for a set of model parameters the only variable allowed to change was the phase of the secondary.

## 4. DISCUSSION

### 4.1. *Synthetic Line Profiles In and Out of Eclipse*

While use of the standard temperature profile given by equation (2) resulted in good fits of synthetic line profiles to observations for a number of CVs (Papers I and II), no satisfactory fits to the C IV line of V347 Pup were found for realistic parameters. The fundamental problem is the high white dwarf mass, the associated small $R_{\rm WD}$, the high assumed mass-accretion rate, and their effect on the maximum disk temperature, which from equation (2) scales like $T_{\rm eff,max} \propto M_{\rm WD}^{1/4} \dot{M}_a^{1/4} R_{\rm WD}^{-3/4}$. For the adopted mass and radius of the white dwarf in V347 Pup, $T_{\rm eff,max} \approx 110{,}000$ °K for $\dot{M}_a = 10^{-8}$ $M_\odot$ yr$^{-1}$. Such high temperatures in the inner disk results in a high ionization state of the LDW unless its mass-loss rate, $\dot{M}_w$, is nearly equal to the accretion rate through the disk. We note that as $\dot{M}_w$ approaches $\dot{M}_a$ little of the originally accreted mass reaches the inner disk and the temperature is significantly reduced. Such a high mass-loss solution based upon the standard temperature profile suffers from several problems which make it unrealistic.

First, the mass-loss rate is unrealistically high for a LDW. Since the asymptotic wind velocity is assumed to scale with the local escape velocity, the outflowing kinetic energy of this wind is a large fraction of the total accretion disk continuum luminosity. However, LDWs are very inefficient at converting the driving radiation energy into wind kinetic energy. This stems from the fact that the photon scattering results in little energy loss even though the momentum transfer may be strong. If the driving force behind the wind is magnetic torque such a large mass-loss rate could in principle be possible.

More significantly, the temperature from equation (2) is inconsistent in that the the kinetic energy transfered from the radiation field to the wind was ignored. We correct for the wind kinetic energy using a simple energy model where we uniformly scale $T_{eff}$ from equation (2) at all radii by $(1 - E_{KE}/E_G)^{1/4}$, where

$$E_{KE} = \int_{R_{RW}}^{R_d} \frac{\dot{M}_w(r) v_\infty^2}{2} 2\pi r dr \tag{3}$$

is the rate at which kinetic energy is carried away by the wind and

$$E_G = \int_{R_{WD}}^{R_d} \frac{3GM_{WD}\dot{M}_a(r)}{4\pi\sigma r^3} \left[1 - \left(\frac{R_{WD}}{r}\right)^{1/2}\right] 2\pi r dr \tag{4}$$

is the rate at which gravitational energy is released. Both integrals are evaluated over the accretion disk.

With this reduction in the disk temperature we been able to reduced the wind-to-accretion mass-loss ratio to 40%. For the model parameters given in Table 1 we plot the temperature and line profiles in Figure 1. In this and all subsequent plots showing the *IUE* spectra, we have shifted the in eclipse observed line profile by 0.75 Å to align the peak intensity with the synthetic spectrum. This is within the uncertainty of *IUE* data at 1550 Å. To allow for instrumental broadening our synthetic profiles were smoothed with a Gaussian of FWHM $\approx 4.5$ Å. We have ploted in Figure 1b the difference between the synthetic line profiles out of eclipse and in eclipse. The double peaked profile of the differenced spectrum is a result of the narrowing of the in eclipse profile relative to the out of eclipse profile.

Although we were able to fit the observed spectra quite well using the disk temperature from equation (2) corrected for energy conservation, we feel that even $\dot{M}_w/\dot{M}_a \approx 0.4$ is unacceptably high for a LDW. A further decrease in the mass-loss rate for the given accretion rate requires a reduction of the photoionization flux from the inner disk. Holding all other parameters fixed at the values given in Table 1, the wind mass-loss rate could be reduced to $\dot{M}_w = 1 \times 10^{-9}$ $M_\odot$ yr$^{-1}$ $= 0.1\,\dot{M}_a$ if the maximum disk temperature was limited to $T_{\rm max} = 50{,}000$ °K. The resulting line profiles are shown in Figure 2a. In this modified fiducial model, the maximum disk temperature, and consequently the inner boundary of the wind, was limited by holding the temperature fixed at 50,000 °K interior to $r_0 = 6.8\,R_{\rm WD}$ (see Figure 2b). To demonstrate the relative effect of $T_{\rm max}$ on the lines in and out of eclipse, we calculated the line profiles for the cutoff temperatures of 60,000 and 70,000 °K as well. As seen in Figure 2a, the relative intensities of in and out of eclipse lines become progressively smaller with increasing $T_{\rm max}$. The simple explanation to this effect is that with increasing ionization in the wind, the integral emission for the line is weighted towards more compact regions around the disk which are successfully eclipsed by the secondary.

Flattening of the inner disk brightness temperature was observed recently for half of the nova-like CVs eclipse-mapped by Rutten, van Paradijs & Tinbergen (1992). The possible explanations suggested by these authors for this phenomena are: (1) a radially extended boundary layer between the white dwarf and the disk; (2) non-radiative cooling of the central part of the disk, e.g., by a wind; and (3) the disruption of the inner disk by the white dwarf's magnetic field. The latter possibility implies that the system is an intermediate polar. Based on the shapes of the profiles of optical emission lines, Williams (1989) concludes that many novae and nova-like variables are intermediate polars. Because of its strong He II $\lambda$4686 emission and the high value of the X-ray to optical flux ratio inferred from the identification of V347 Pup with 4U 0608−49, Buckley et al. (1990) speculated that V347 Pup itself might be an intermediate polar (however, see Paper III). We note that several nova-like CVs have been reclassified or are suspected of being magnetic variables, e.g., nova-like 1H1752+081 (Silber et al. 1994). It is interesting that observations of the bright nova-like IX Vel using the *Hopkins Ultraviolet Telescope* have revealed a similar trend: theoretical continuum spectrum for this object is far too blue and gives too high a flux in UV (Long et al. 1994). Alternatively, the inner disk can be evaporated by a coronal siphon flow (Meyer & Meyer-Hofmeister 1994), or be susceptible to other instabilities.

To investigate the possibility that the maximum disk temperature in V347 Pup is limited to $\approx 50{,}000$ °K due to the disruption of the inner disk by evaporation or by the white dwarf's putative magnetic field, we calculated a set of models in which the disk was truncated at the inner radius of $r_0 = 6.8\,R_{\rm WD}$ where $T_{\rm eff}(r_0) = 50{,}000$ °K. This is a crude approximation to what actually happens in a magnetic disk (e.g., Mauche et al. 1990), but is sufficient for our purposes. Figure 3 shows that there is not much difference between models that have their temperature limited to 50,000 $^o$K and those where the inner disk was completely cut-off above this temperature. For that reason, we limit our subsequent discussion to models with a flat inner disk temperature profile.

For the model parameters which reproduce well the line spectra in and out of eclipse, the LDW is very optically thick in the C IV line. Because the radiation transport is insensitive to local density and ionization changes in the optically thick part of the outflow, the line profiles are insensitive to variations in most of the model parameters. Moreover, due to the large optical depth in the line, the C IV emission which otherwise would appear as double-peaked at this inclination due to rotational broadening is single-peaked because emission from the extended wind fills in between the double peaks. The angle-averaged Sobolev escape probability for the ground state of this line is shown in Figure 4. It becomes 1/2 only at $\approx 3a$ above the disk — comparable to the disk and secondary sizes in this system, before the wind reaches its terminal velocity. The ionization contours for our generic model are given in Figure 5, and reflect the growing exposure of wind to the radiation coming from the central disk. The ionization flux depends strongly on the limb darkening which provides partial shielding close to the disk. The ground state ionization fraction of C IV peaks at around 80%, and drops to 3% at very large distances from the white dwarf.

In order to test the sensitivity of the generic model to different parameters, we have varied them, typically within a factor of two in both directions. Figure 6 shows the effect of changing $\dot{M}_w$ on the line profiles. It is obvious immediately that this type of perturbation affects the emission line in eclipse much more than that out of eclipse. Decreasing $\dot{M}_w$ by a factor of 2 reduces the density in the LDW, increases its ionization and results in reduced emission in the line's core and vice versa. We find that increasing/decreasing $\dot{M}_a$ has an effect similar to decreasing/increasing $\dot{M}_w$. The ratio $\dot{M}_w/\dot{M}_a$ provides a robust characteristic of the wind in the sense that the line profiles in and out of eclipse stay nearly the same as long the this ratio is preserved. We also note that the line profiles are quite insensitive to the choice of inclination angle for $i \gtrsim 85^o$. For smaller inclination angles the continuum emission becomes significantly larger than that observed by *IUE*.

The introduction of rotation in the wind strongly modifies the line profile through rotational Doppler broadening. As shown in Figure 7 (see also Fig. 19a of Paper III), there is a dramatic change in the line profile when the rotation of the wind is switched off. The out of eclipse line narrows and is now in clear disagreement with the observed profile. Because the secondary covers the central part of the wind which rotates most rapidly, the eclipsed line is not significantly affected. This underlines the importance of rotation in shaping the emission lines from disk winds.

### 4.2. *Time-Resoved Eclipse in V347 Pup: Numerical Model*

The present synthetic spectral fitting of V347 Pup is restricted to two orbital phases only as *IUE* requires exposures of typically 20 min. With its larger mirror and higher efficiency detectors, the *HST* is capable of time-resolving the eclipse in CVs. We therefore provide a numerical mapping of time-resolved eclipse in V347 Pup, i.e., synthetic eclipse for the C IV line, in anticipation that the predicted behavior will be confirmed in future *HST* observations. Figure 8 displays the line intensity evolution as the eclipse progresses between the phases $\phi = -45^o$ (entering the eclipse) to $\phi = +45^o$ (leaving the eclipse), with the maximal eclipse at $\phi = 0^o$. The evolution of the synthetic line profiles with eclipse phase are shown in Figure 9.

In our model the disk, wind, and secondary rotations are all in the positive $\phi$ direction. As the eclipse begins, the secondary first blocks emission from the disk and wind rotating towards the observer. The blue core of the line, which originates in the slowly rotating wind at large distances from the rotation axis, is first affected ($\phi \approx -26^o$). The eclipse propagates towards the blue wing of the line (wind rapidly rotating towards the observer and close to the axis; $\phi \approx -20^o$), and then to the red wing ($\phi \approx -17^o$) as the wind from the inner disk rotating away from the observer is eclipsed. Finally, the red core decreases as one approaches the full eclipse at $\phi = 0^o$ and the more slowly rotating components of the wind rotating away from the observer are covered. The decrease in the red component of the line profile is abrupt ($\Delta\phi \approx 0.5^o$) compared to the blue component ($\Delta\phi \approx 2^o$) — an effect which can be clearly observed with the *HST*. This behavior is due to the fact that changes in the blue line component come from obscuration first of the slowly rotating portion of the wind followed by the rapidly rotating region. For the red line component the order is reversed, with the rapidly rotating wind being first eclipsed and then the slowly rotating region. As one moves out of eclipse the blue core first recovers, followed by the blue wings, the red wings and then the red core. Again, the changes in the red line component is more abrupt that that for the blue due to the asymmetry of the eclipsing of high and low rotational components of the wind.

During the eclipse, the blue wing taken at 10% intensity level is reduced from -2,220 to -1,640 km $s^{-1}$ and the red wing — from +1,830 to +1,250 km $s^{-1}$. Hence, the blue asymmetry of the line, i.e., that of the centroid, is preserved during the eclipse, except for a narrow range $\phi \simeq -20^o$ to $-18^o$, when the line centroid appears redshifted, which simply reflects the fact that the eclipse propagates in the general direction of rotation in the system.

## 5. CONCLUSIONS

High mass-transfer rates inferred for nova-like CVs stabilize their accretion disks, thermally and dynamically. This simplifies considerably our analysis of UV line spectra in these objects and partially justifies our assumption of a steady state winds from disks. Because of its almost edge-on equatorial plane, the eclipsing nova-like CV V347 Pup appears to be a prototype system for studying the effects of rotation on UV emission line shapes.

Using 3D kinematic and radiative transfer models for the disk wind, we have obtained good fits to the in and out of the eclipse C IV line by fixing all the parameters of the system and allowing only for changing phase of the secondary. The narrowing of the line profiles during the eclipse can be explained by secondary covering the innermost part of the wind. The obscured part located close to the rotation axis of the disk is contributing mostly to the rotational broadening of the line. Narrowing of the C IV during the eclipse was also observed by Mason et al. 1995 for UX UMa using the *HST*.

V347 Pup has an unusually massive white dwarf which results in accretion disk extending to a very small distance from the rotation axis and, consequently, emitting exceedingly large flux in UV. To prevent over-ionizing the wind, the wind density must be quite high. Reasonable fits to in and out of the eclipse line have been obtained for $\dot{M}_w/\dot{M}_a \gtrsim 0.4$ using the standard blackbody disk temperature profile corrected for mass and energy conservation. We note, that such high mass-loss rates would be unacceptable for the line-driven wind from a disk, but it would be at least possible for a hydromagnetically-driven wind. In order to minimize the mass-loss to accretion rate ratio in V347 Pup, we have produced a series of numerical experiments by varying the inner disk temperature profile. Successful fits for $\dot{M}_w/\dot{M}_a \approx 0.1$ were obtained by limiting the maximal temperature in the disk to 50,000 °K. This was done either by leveling off the temperature profile in the disk in agreement with the recent observations of the nova-like CVs, or by the cutting off the innermost disk around 6.8 white dwarf radii either by disk evaporation or by the magnetic field originating on the white dwarf. In this latter case, V347 Pup would be an intermediate polar. Much lower ratios of $\dot{M}_w/\dot{M}_a$ can be achieved by further limiting the maximal temperature in the disk to around 30,000-40,000 °K. We also emphasize that our results are weakly dependent upon the ratio $\dot{M}_w/\dot{M}_a$ and are therefore compatible with even lower accretion rates than $10^{-8}$ $M_\odot$ $yr^{-1}$ adopted in this work.

High-speed spectrophotometry by the *HST* should be capable of time-resolving eclipse in V347 Pup and most of the other high inclination CVs. We thus provide a numerical model for such a synthetic eclipse in this object. For disk winds we predict a narrowing of the line profile during eclipse. This narrowing occurs abruptly, first in the blue and then in the red components of the line.

We conclude that rotation, which appears to be the major difference between accretion disk winds and those centered on the white dwarfs, is also necessary to explain the behavior of UV lines in V347 Pup, and other similar systems, in and out of the eclipse. For the first time rotation has been positively identified as a necessary component in the CV winds. This was accomplished using our fully 3D modeling of radiative transfer there. Our results confirm that the accretion disk and not the white dwarf is the source of wind in these systems.

ACKNOWLEDGEMENTS: We thank Rick Hessman, John Raymond and Rainer Wehrse for stimulating discussions. This work was supported in part by NASA grant NAGW-3839 and NASA/EPSCoR grant to the University of Kentucky. P.V. and C.W.M.'s contribution to this work was performed under the auspices of the U.S. Department of Energy by Lawrence Livermore National Laboratory under contract No. W-7405-Eng-48.

**TABLE 1**
Model Parameters

| | |
|---|---|
| $i$ | 89° |
| $M_{\rm WD}$ | 1.2 $M_\odot$ |
| $R_{\rm WD}$ | $4 \times 10^8$ cm |
| $T_{\rm WD}$ | 40,000 °K |
| $T_{\rm BL}$ | 10,000 °K |
| $\dot{M}_a$ | $1 \times 10^{-8}$ $M_\odot$ yr$^{-1}$ |
| $\theta_{\rm min};\ \theta_{\rm max}$ | 5°; 30° |
| $R_d$ | 100 $R_{\rm WD}$ |
| $T_{\rm min};\ T_{\rm max}$ | 20,000; 50,000 °K |
| $r_{\rm min};\ r_{\rm max}$ | 6.8 $R_{\rm WD}$; 50 $R_{\rm WD}$ |
| $R_v$ | 300 $R_{\rm WD}$ |
| $v_\infty;\ \alpha$ | 3 $v_{\rm esc}$; 1.5 |
| $R_s$ | $4 \times 10^8$ cm |
| $a$ | $1.3 \times 10^{11}$ cm |

## FIGURE CAPTIONS

Fig. 1a.—Disk temperature and continuum intensity for the $T_{eff}$ from equation (2) scaled by $(1 - E_{KE}/E_G)^{1/4}$. Model parameters are given in Table 1. The blackbody intensity for temperature $T_{\rm eff}(r_0)$ has been weighted by $2\pi r_0$ to give the intensity per unit length in disk radius $r_0$.

Fig. 1b.—Line profiles in and out of eclipse for the model described in Figure 1a. Observed profiles are represented by dot-dashed lines; synthetic profiles by solid lines. The lower curves give the difference between spectra out of eclipse and in eclipse.

Fig. 2a.—Line profiles in and out of eclipse for the modified fiducial model (Table 1) with $\dot{M}_w \approx 10^{-9}$ M$_\odot$ yr$^{-1}$ for our constant inner disk temperature model. Synthetic line profiles for a maximum inner temperature of 50,000 °K, 60,000 °K, and 70,000 °K are shown. Solid curves give synthetic profiles. The dot-dashed lines represent observed profiles.

Fig. 2b.—Disk temperature and continuum intensity for conditions given in Figure 2a. The blackbody intensity for temperature $T_{\rm eff}(r_0)$ has been weighted by $2\pi r_0$ to give the intensity per unit length in disk radius $r_0$.

Fig. 3.—Comparison of line profiles produced by the modified fiducial model with $\dot{M}_w \approx 10^{-9}$ M$_\odot$ yr$^{-1}$ with constant $T_{\rm eff}$ = 50,000 °K in the inner disk (*dot-dashed curves*) and the same model with a central hole of radius $r_0 = 6.8 R_{\rm WD}$ where $T_{\rm eff}(r_0)$ = 50,000 °K (*solid curves*).

Fig. 4.—Sobolev escape probability from the wind for the ground state of C IV for the modified fiducial model with $\dot{M}_w \approx 10^{-9}$ M$_\odot$ yr$^{-1}$.

Fig. 5.—C IV ionization contours for the modified fiducial model with $\dot{M}_w \approx 10^{-9}$ M$_\odot$ yr$^{-1}$.

Fig. 6.—Effect of changing $\dot{M}_w$ on the C IV line profile.

Fig. 7.—Synthetic line profiles (solid lines) with wind rotation *ignored* out and in the eclipse superimposed on the observed line profiles.

Fig. 8.—Synthetic eclipse in V347 Pup. The arrows labeled by a–g correspond to various eclipse phases (explicit C IV line profiles are given in Fig. 9). Eclipse proceeds from $a$ (initial phase) to $g$ (final phase). Maximal eclipse phase is $d$. Both solid and dashed lines are iso-intensity curves which differ by 5% from each other. They are labeled as fractions of the C IV line peak intensity. As the eclipse progresses, the iso-intensity curves are blue- and red-shifted, respectively, towards the line peak, i.e. the line becomes much narrower and slightly dimmer.

Fig. 9.—Synthetic eclipse in V347 Pup. Variation of C IV synthetic line profile with orbital phase. The curves marked a–d (upper figure) correspond to the eclipse phases $\phi = -30^o$, $-20^o$, $-10^o$, $0^o$ (see Fig. 8). Curves marked d–g (lower figure) correspond to $\phi = 0^o$, $10^o$, $20^o$, and $30^o$, respectively.